# Quantum Conductance and Electronic Properties of Lower Diamondoid Molecules and Derivatives

***Yong Xue*** *

*Department of Physics*

*University of Illinois at Chicago*

*Chicago, IL 60607-7052, USA*

***G. Ali Mansoori*** #

*Departments of BioEngineering, Chemical Engineering and Physics*

*University of Illinois at Chicago*

Chicago, IL 60607-7052, USA

## Abstract

Diamondoids and their derivatives have found major applications as templates and as molecular building blocks in nanotechnology. Applying *ab initio* method, we calculated the quantum conductance and the essential electronic properties of two lower diamondoids (adamantane and diamantane) and three of their important derivatives (amantadine, memantine and rimantadine). We also studies two artificial molecules that are built by substituting one hydrogen ion with one sodium ion in both adamantane and diamantane molecules. Most of our results are based on an infinite Au two-probe system constructed by ATK and VNL software, which comprise TRANSTA-C package. By changing various system structures and molecule orientations in linear Au and 2×2 Au probe systems, we found that although the conductance of adamantane and diamantane are very small, the derivatives of the lower diamondoids have considerable conductance at specific orientations and also showed interesting electronic properties. The quantum conductance of such molecules will change significantly by changing the orientations of the molecules, which approves that residues like nitrogen and sodium atoms have great effects on the conductance and electronic properties of single molecule. There are obvious peaks near Fermi energy in the transmission spectrums of artificial molecules, indicating the plateaus in *I-V* characteristics of such molecules.

---

* E-mail: *yxue4@uic.edu* # Corresponding author, E-mail: *mansoori@uic.edu*

## I. Introduction

Diamondoid molecules are cage-like, ultra-stable, saturated hydrocarbons. These molecules are ringed compounds, which have a diamond-like structure consisting of a number of six-member carbon rings fused together (see Table I). They are called "diamondoids" because their carbon-carbon framework constitutes the fundamental repeating unit in the diamond lattice structure. Due to their six or more linking groups, they have found major applications as templates and as molecular building blocks in nanotechnology, polymer synthesis, drug delivery, drug targeting, DNA-directed assembly, DNA-amino acid nanostructure formation, and host-guest chemistry. Diamondoid derivatives have excellent potential for building diamondoid-based nano-electro-mechanical systems (NEMS) and other nanoscale logic units. [1-5]

In this paper, we report our *ab initio* studies of the quantum conductance and electronic properties of lower diamondoid molecules and some of their important derivatives. Specifically seven molecules two of which are lower diamondoids and the rest are important diamondoid derivatives are chosen. We classified them into three groups as shown in Table I: Group 1: Adamantane (*ADM*), Diamantane (*DIM*) the lowest two diamondoids. Group 2: Memantine, Rimantadine and Amantadine, the three derivatives of adamantane which have found tremendous amount of medical applications as antiviral agents[5,6]. Group 3: ADM•Na, DIM•Na, the two artificial molecules, substituting one hydrogen ion in adamantane and diamantane with a sodium ion. The latter group has potential applications in NEMS.

We studied the quantum conductance and electronic properties of the three groups and analyzed our results. One important goal of molecular electronics is to find functional molecular structures that can be used as the logic units with which we can build nanoscale integral circuits. There have been many studies in the use of single molecules as "functional electronic devices" and to find the transportation properties of different molecules and molecular building blocks[7-9]. Instead of studying monolayer molecules in previous cases, however, we studied diamondoid molecules and their derivatives, an important group of organic molecules which have 3-Dimensional structures and more possibilities as molecular building blocks for nanotechnology [2,5]

## II. Computational Procedure

In our calculations, we used the ATK (Atomistix ToolKit) software package (*www.atomistix.com*) and the VNL (Virtual Nanolab) software package (*www.virtualnanolab.com*). Atomistix ToolKit (ATK) whose predecessor is TranSIESTA-C package[10,11] is an *ab initio* electronic structure program capable of simulating and modeling electrical properties of nanostructured systems coupled to semi-infinite electrodes. Non-equilibrium Green's functions (NEGF) and density functional theory (DFT) are combined in the software and the entire system was treated self-consistently under finite bias conditions[10]. The Virtual Nanolab (VNL) software package is based on ATK and gives access to atomic-scale modeling techniques with a graphical interface for simulation and analysis of the atomic scale properties of nanoscale devices.

In most cases of calculations reported here, we used LDA-PZ, which is the local density approximation (LDA) with the Perdew-Zunger (PZ) parametrization[12] of the correlation energy of a homogeneous electron gas calculated by Ceperly-Alder[13]. In these *ab initio* electronic structure computations, we used the Double Zeta Polarization (DZP) basis set. It is important to choose a basis set large enough to give a good description of the molecular wave function of diamondoids. Double Zeta Polarization (DZP) basis set is one such basis set. A double-zeta basis set for hydrogen has two functions, and a true double-zeta basis set for carbon would have ten functions. Additional flexibility is built in by adding higher-angular momentum basis functions. Since the highest angular momentum orbital for carbon is a *p* orbital, the "polarization" of the atom can be described by adding a set of *d* functions. In our *Ab initio* calculations, the optimal value of the mesh cut-off was found to be 100 Ry, and the optimum *k*-point grid mesh number in the Z direction was found to be 100.

In order to calculate the electronic properties of nanoscale systems, a two-probe simulation system was created. This system consists of two semi-infinite electrodes and a scattering intermediate region which includes several layers of molecules from each electrode. In the practical calculations, only the central region (region between the two electrodes) and the scattering regions are considered, and the bigger the scattering region, the more accurate the calculation will be but also more computation efforts needed.

We constructed two types of electrodes: first type is semi-infinite linear gold (Au) chains and second one is comprised of Au (100) surface in a 2×2 unit cell, as shown in Figure 1. We choose gold as the electrodes since it is more practical and promising as monatomic nanowire. The constant bond length of Au atom was optimized by ATK. Then by performing separate calculations for the scattering region as well as the central region, where the diamondoid molecule is located, followed by an intelligent recombination of the two subsystems, the quantum transmission spectrums and conductance of diamondoids can be calculated.

The quantum conductance, *G*, is calculated by the following equation [9],

$$G=G_0 \cdot T(E, V_b), \qquad (1)$$

where $G_0 = 2e^2/h = 77.5\mu S$.

$T(E, V_b)$ is the transmission probability for electrons incident at an energy $E$ through a device under a potential bias $V_b$.

The current through the scattering intermediate region is determined by the quantum-mechanical probability for electrons to tunnel through the diamondoid molecule from one electrode to the other. This current is calculated using the Landauer formula which expresses the conductance of a system at T=0 in terms of the quantum mechanical transmission coefficients [9],

$$I = \int_{\mu_L}^{\mu_R} T(E, V_b) dE \,, \qquad (2)$$

Where $\mu_L$ and $\mu_R$ are the left- and right-side metallic reservoirs electrochemical potentials ($\mu_L > \mu_R$),

$\mu_{L/R} = \pm eV_b$ and $T(E, V_b)$ is the transmission probability for electrons incident at an energy $E$ through a device under a potential bias $V_b$. When $V_b > 0$ it means that positive charge transport from the right electrode. The Landauer equation based on the Green's function method relates the elastic conductance of a junction to the probability that an electron with energy $E$ injected in one electrode will be transmitted to another electrode through a scattering region which in our case is the diamondoid molecule.

Using the above-mentioned procedure, we calculated the energy levels, transmission spectrums and conductance of the seven chosen molecules. We also calculated current-voltage (*I*-*V*) and conductivity-voltage (*G*-*V*) characteristics of some of those molecules.

In this paper, we have focused on the exhibition of the results we obtained and the prospective applications in the field of nanotechnology. As a simulation of real experiments, geometrical optimizations using DFT method is necessary. However, since diamondoids are very stable and stiff molecules therefore, the geometrical structures of these molecules would not, and thus the conductance would not, be affected. For these reasons, we do not need to perform geometrical optimizations. We checked by optimizing adamantane and diamantane with two gold (Au) atoms considered as the two-tips of the electrodes. In both cases, the geometrical structures of these diamondoids did not change significantly. Therefore, we could conclude that due to the stiffness of diamondoids, the electronic properties would not be affected by the two tips that are used as the electrodes. Although we did not optimize the geometrical structures of molecules in Group 1 and Group 2, we indeed optimized geometrical structures of some two-probe systems which include molecules and electrodes using ATK, and also we did geometrical optimization of the molecules in Group 3. We would like to discuss the results in the following sections.

## III. Computational Results

### A. HOMO and LUMO

First, we calculated the highest occupied molecular orbital (HOMO) and the lowest unoccupied molecular orbital (LUMO) of all the seven molecules as reported in Table II. The smaller the band gap (the difference between the energies of the HOMO and LUMO) is indicative of the fact that the more easily a molecule can be excited. Our computational results of the first group are generally in agreement with previous calculations reported in the literature [8]. The small deviation between our calculations and the literature data are due to different basis and approximation methods applied. In this paper, however, we focused on the changes of HOMO-LUMO gap from Group 1 to Group 3. From Table II , we see that Group 1 (the lower diamondoids) have the largest band gap, which proves that diamondoids are electrical insulators. The three derivatives in Group2 have smaller band gaps, which indicate that those molecules could be electrical semiconductors. The results also indicate that the -$NH_2$ group is electron donating. This is consistent with the fact that -$NH_2$ has a pair of non-bonded electrons. Group 3 shows not only smaller band gaps than Groups 1 and 2 which indicate high conductance, but also interesting electronic behaviors, such as both the HOMO and LUMO are below Fermi energy which is an indicator of the metallic properties of those molecules. We would also like to mention that the HOMO and LUMO of molecules are independent of those of the two-probe systems. Following these results and directions, we attempt to study the quantum conductance of different groups at different orientations.

### B. Quantum conductance and transmission spectrums at zero bias

In order to study the conductance of diamondoid molecules and their derivatives and to simulate the experimental conditions, the geometrical optimizations of two-probe systems are performed. Applying DFT method, we first optimized the distances between the linear Au electrodes and the Adamantane and Diamantane molecules and the optimal values are 9.06 Å and 11.00 Å respectively. We did not optimize the geometries of derivatives of Diamondoids with electrodes, however, since: 1. These molecules are not geometrically symmetrical, and if we optimize the structures of the whole systems including Au electrodes, the orientations of these molecules will change which is not what we expected. 2. Our purpose is to point out the potential of building nano scale electronic devices using the diamondoids and their derivatives, not the structures of the devices which could be manipulated in order to obtain useful conductance. Although the geometrical optimization of all the systems are not such necessary, we considered the optimized distances between the electrodes and Adamantane and Diamantane molecules i.e. 9.06 Å and 11.00 Å when choosing the distance between the electrodes and molecules inside, i.e. the central region width .

Based on the geometry considerations above, we first calculated the quantum conductance and transmission spectrums of all our systems at zero bias. All the data are reported in Table III, Figure 2 and Figure 3.

#### 1. Adamantane and Diamantane

To check our method, we began the calculations with adamantane. As expected, the conductance of adamantane as reported in Table III is quite small. From the transmission spectrum reported in Figures 2(a) and 3(a) we can conclude that adamantane is an electrical insulator. It seems that our method is pragmatic. For diamantane, we obtained similar results (see Table III and Figures 2(a) and 3(a)). However, since the HOMO-LUMO gap of diamantane (Table II) is smaller, we obtained a higher conductance. As mentioned above, we used both Au linear chains and Au (100) 2×2 unit cells as electrodes, and from the results, we found that molecules confined in Au 2×2 electrodes have higher conductance than those confined in Au linear chains. Even with the same central region widths as the linear Au electrodes cases ( 9.06 Å and 11.00 Å) the conductance of Adamantane and Diamantane in Au 2×2 electrodes case (1.9375$\mu S$ and 0.418465$\mu S$) are still greater than those in linear case (0.042$\mu S$ and 0.137$\mu S$ ). An explanation could be that the diamondoids are 3-dimensional structures, and when electrodes are also with 3-dimensional structures, the entire systems could have more conductive channels and thus have higher conductance than when electrodes are linear. Nevertheless, this does not mean when we construct molecular electronic structures we have to use 3-dimensional electrodes, we will further discuss this in the next section.

### 2. Memantine, Rimantadine and Amantadine

When we calculated the conductance of this antiviral group of diamondoid derivatives, we chose various orientations by rotating the molecules between the two electrodes. The two reasons for these rotations are: first, molecules in this group are functionalized and their symmetries are different; second, we wanted to detect the orientations of these molecules that are electronically better which may have applications for electron transfer or electron capture in their interactions with bio-molecules. From this part on, the geometrical optimizations of the entire systems was not necessary any more since we were studying the effect of orientation of molecules on their electronic properties. Since the non-optimized structures might influence the accuracy of our calculations, our results for different orientations and various molecules should be used for comparison purposes only.

The conductance obtained for memantine and the central region widths at three different orientations are reported in Table III. According to this table, in linear electrode case memantine conductance changes slightly when we rotate the molecule. However, in the case of 3-Dimensional electrode we can see significant changes in memantine conductivity with orientation change. When the nitrogen (-N) ion is close to an electrode (Orientation 2), the conductance is the largest.

From the transmission spectrums, Figure 2(b) and Figure 3(b), we can see that the LUMO resonance has more contributions to the conductance of memantine. The above two results also proved that the -$NH_2$ group has the function of n-donor.

Then we studied amantadine at two different orientations as reported in Table III. From Table III, we can see that the conductance changes significantly with different types electrodes. The transmission spectrums of amantadine, as reported in Figures 2(b) and 3(b), are quite similar in trend to that of memantine.

We performed similar calculations for rimantadine, the last member of this group, at two different orientations as reported in Table III. The nitrogen ion in rimantadine is not directly connected to the adamantane cage, which makes it a bit different from memantine and amantadine. As it is reported in Table III the conductances are small in Orientation 1 of this molecule for both kinds of electrodes. However, we obtain high conductances in orientation 2. The Orientation 2 conductances of this molecule are the highest among all the three molecules in all orientations. The transmission spectrums of rimantadine, as reported in Figures 2(b) and 3(b) are quite distinct from those of memantine and amantadine in different ways for linear and for 2×2 electrodes.

### 3. ADM•Na and DIM•Na

We studied the quantum conductance of ADM•Na and DIM•Na, two artificial diamondoids derivatives as well as the single sodium (Na) atom. As it is reported in Table III, it is obvious that the conductances of these artificial molecules are generally higher than those of the lower diamondoids and derivatives reported in Table III. It is also shown that the conductance is a strong function of the position of the Na ion in the molecule: the closer the Na ion to an electrode, the greater the conductance. The difference is large for the two orientation because from the transmission spectrum the rotation of ADM•Na molecule actually changes the Fermi energy i.e. when the Na ion is close to the electrode, the Fermi energy is close to the peak of the transmission spectrum, while the Na is far from the electrode, the Fermi energy is above the peak. From this result, we can predict that if we can find a method (electric or magnetic field, or some mechanical force, etc) to rotate the ADM•Na molecule, then we can control the conductance of this nanostructure and we could build a nanoscale transistor.

We also studied DIM•Na and we obtained similar results to ADM•Na. Moreover, we found another interesting result for this molecule: a sharp transmission peak is shown to be close to the Fermi energy, see Figure 2(c). We can predict that as the bias increases, the current would not increase and there should be a peak or plateau on the *I-V* characteristics graph, as discussed in the next section. Moreover, in all the Au 2×2 electrode cases, conductance is greater than those in the three linear cases, especially, in Orientation 1. Another interesting result is that in Orientation 2 the conductance of DIM•Na with Au 2×2 unit cell is much greater than ADM•Na with similar structure in which the conductance is almost zero. In addition, this conductance is much higher than the metallic single sodium atom conductance. This molecular orientation is predicted to be metallic representing a substantial change in electronic properties relative to the non-conducting diamondoids. Similar effects were observed in nanotubes and fullerene doped with alkali metals [14-16].

At the end of this section, we would like to compare the transmission spectrums which demonstrate the electronic properties of our systems. From Figure 2 (a) and Figure 3 (a) we can see that Group 1 is an insulator group, especially, adamantane has very low conductance. According to Figures 2(b) and 3(b) although the conductance of Group 2 is not high but from the transmission spectrum [see Figure 2 (b) and Figure 3 (b)] we find that conductance of Group 2 can approach to $1G_0 = 77.5\mu S$. Therefore, we predict that Group 2 has the potential to build molecular electronic devices and the results in Table III may be used to explain the medical applications of the molecules in this group. From Figure 2(c) and Figure 3(c), we can see the most active transmission spectrums, which show the high conductance of two artificial molecules ADM•Na and DIM•Na.

## IV. *I-V and G-V* characteristics

In the last section of this paper, we applied bias potentials on electrodes and we studied the *I-V* and *G-V* characteristics of ADM•Na and DIM•Na molecules. Such information can be useful to build nanoscale active devices such as diodes or transistors. By studying the effects of applying various steady voltages or currents; we would find a desired mode of operation. As an example, we report horizontal orientation of these two molecules when placed in between Au linear Electrodes as reported by Figure 4. According to Figure 4, there are plateaus in the *I-V* graph between 0.5$V$ and 0.9$V$ for both cases of ADM•Na and DIM•Na molecules and the currents come to maximums of 11.90$\mu A$ and 11.35$\mu A$. Since the current is the integral of transmission over energy as in Eq. (2) and when bias $V_b$ increases transmission $T(E, V_b)$ decreases, therefore the integral would come to a maximum, which is the reason there is a plateau in the *I-V* characteristic graph. Using this *I-V* characteristic, we could build molecular rectifier or surge protector. Furthermore, two characteristics graph are obtained from the linear Au chain since only in the linear electrodes case we found peaks near Fermi energy, which show linear electrodes or atomic nanowire also have possibility to construct nanoscale devices for 3-dimensional molecules.

## V. Discussion

According to our first-principles studies of the three groups of diamondoids and derivatives, we could conclude that when different residues are added to diamondoids, the electronic properties of diamondoids can be changed, such as the HOMO-LUMO gap of molecules, the conductance, etc. We also found that the diamondoid derivatives have the potential as molecular building blocks of nanoscale electronic devices. The distance between the electrodes in the two cases depend on the orientation of the molecule in between the electrodes. Normally, in Au 2×2 Electrodes case the conductance is greater than linear case, since when conductance is calculated, the central region is considered which includes some parts of the electrodes and there are more atoms in the 2×2 case. It is demonstrated that quantum conductance of diamondoid molecules and their derivative molecules change significantly by changing their orientations. The electronic property data and methodology generated is useful for building diamondoid-based nano-electro-mechanical systems (NEMS) and other nanoscale logic units.

**Acknowledgement:** The authors would like to thank Prof. P. Mohazzabi and Dr. H. Ramezani for reading the manuscript and providing us with constructive comments.

## Appendix A: Abbreviations

| | |
|---|---|
| ADM | Adamantane |
| ATK | Atomistix ToolKit |
| DFT | Density functional theory |
| DIM | Diamantane |
| DZP | Double zeta polarization |
| HOMO | highest occupied molecular orbital |
| LDA-PZ | Perdew-Zunger local density approximation |
| LUMO | Lowest unoccupied molecular orbital |
| NEGF | Non-equilibrium Green's functions |
| Ry | Rydberg |
| VNL | Virtual Nanolab |

## Appendix B: Symbols

| | |
|---|---|
| $e$ | Electron charge |
| $E$ | Energy |
| $\mathbf{G}$ | Conductance |
| $\boldsymbol{G_0}$ | Quantum conductance $2e^2/h = 77.5\mu S$ |
| $H$ | Hamiltonian |
| $I$ | Current |
| T | *Transmission* |
| $V_b$ | *Bias potential* |
| $\mu S$ | $10^{-6}$ Siemens |

**Table I.** Molecular formulas and structures of Adamantane, Diamantane, Memantine, Rimantadine, Amantadine, Optimized ADM•Na and Optimized DIM•Na molecules. In these figures blacks represent –C, whites represent –H, Blues represent –N and purples represent –Na.

| **Group 1** | | **Group 2** | | | **Group 3** | |
|---|---|---|---|---|---|---|
| Adamantane | Diamantane | Memantine | Amantadine | Rimantadine | Optimized ADM•Na | Optimized DIM•Na |
| *$C_{10}H_{16}$* | *$C_{14}H_{20}$* | *$C_{12}H_{21}N$* | *$C_{10}H_{17}N$* | *$C_{11}H_{20}N$* | *$C_{10}H_{15}Na$* | *$C_{14}H_{19}Na$* |
| 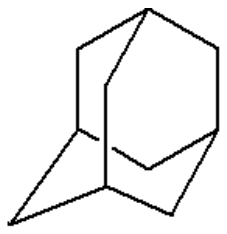 | 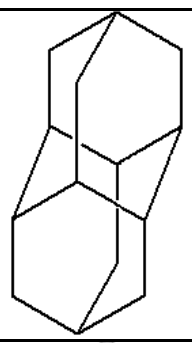 | 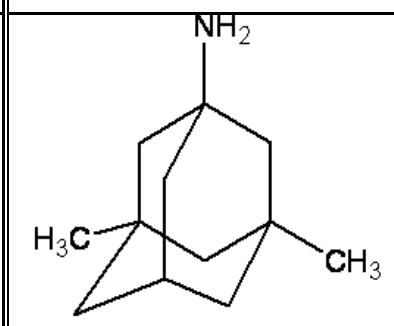  | 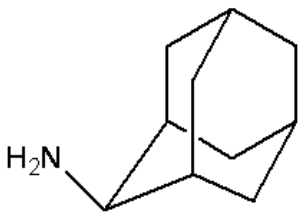  | 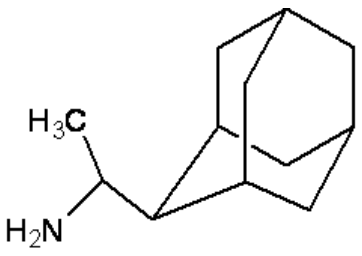  | 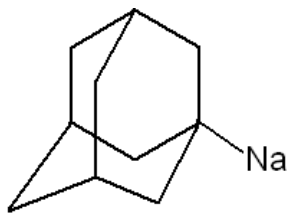  | 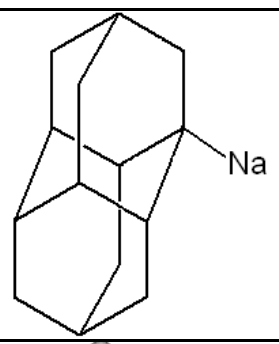  |
| 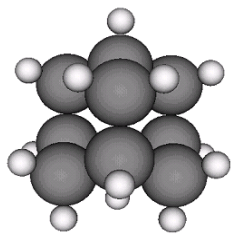 | 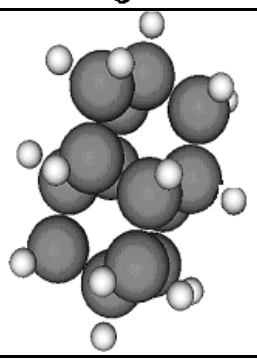 | 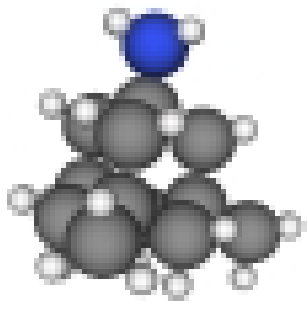 | 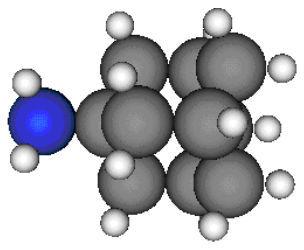 | 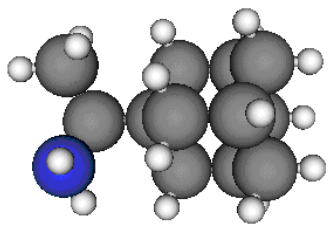 | 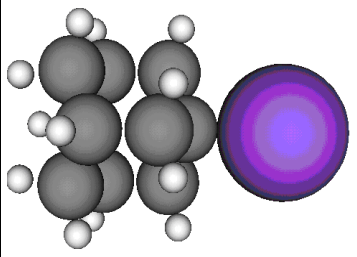 | 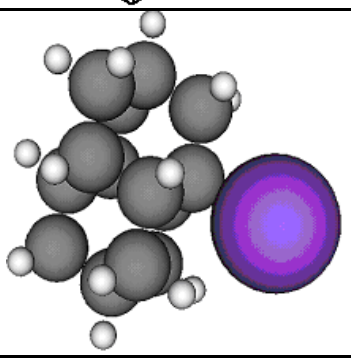 |

**Table II.** HOMO and LUMO of the three groups. (Fermi energy=0 eV)

| | Group 1 | | Group 2 | | | Group 3 | |
|---|---|---|---|---|---|---|---|
| | Adamantane | Diamantane | Memantine | Amantadine | Rimantadine | ADM•Na | DIM•Na |
| HOMO (eV) | -6.554 | -6.226 | -5.023 | -4.956 | -4.975 | -3.078 | -3.435 |
| LUMO(eV) | 1.288 | 1.097 | 1.228 | 1.043 | 0.974 | -1.699 | -1.674 |
| Gap (eV) | 7.842 (7.622*) | 7.323 (7.240*) | 6.251 | 5.999 | 5.949 | 1.379 | 1.761 |

(*) *Calculation by another method* [3].

**Table III.** Quantum conductance ($G$), in [$\mu S$], and Central Region Width, in [Å], of the three diamondoids groups at zero bias and the corresponding orientations

| Molecule | Orientation | **Quantum Conductance** *(Central Region Width)* | |
|---|---|---|---|
| | | Au linear Electrodes 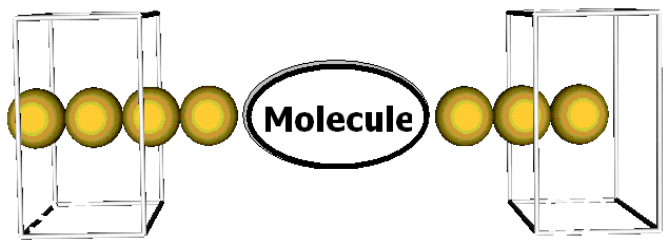  | Au 2x2 Electrodes 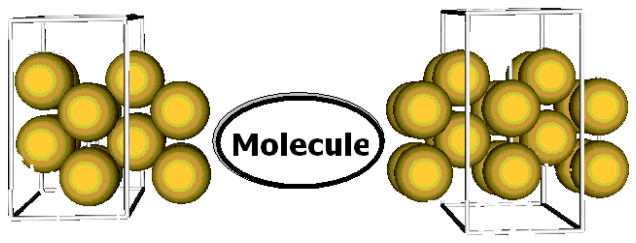  |
| **Adamantane** | 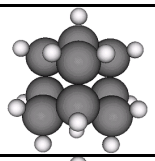 | **0.042** *(9.06)* | **2.393** *(8.15)* |
| **Diamantane** | 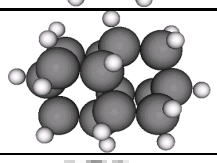 | **0.137** *(11.00)* | **10.785** *(9.76)* |
| **Memantine** (Orientation 1) | 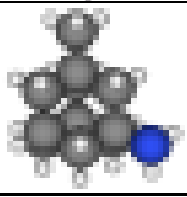 | **0.850** (9.08) | **28.668** *(8.46)* |
| **Memantine** (Orientation 2) | 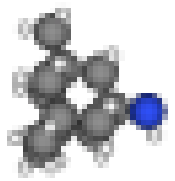 | **1.161** (8.69) | **31.308** (8.65) |
| **Memantine** (Orientation 3) | 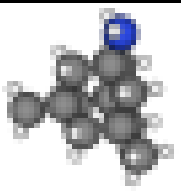 | **0.842** (8.53) | **3.629** (9.64) |
| **Amantadine** (Orientation 1) | 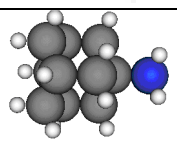 | **0.767** (8.84) | **14.206** (8.68) |
| **Amantadine** (Orientation 2) | 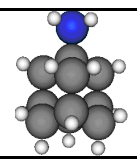 | **8.072** (7.83) | **15.039** (8.10) |
| **Rimantadine** (Orientation 1) | 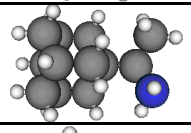 | **0.020** (9.90) | **2.837** (9.85) |
| **Rimantadine** (Orientation 2) | 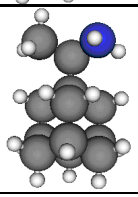 | **11.818** (7.43) | **40.817** (8.32) |
| **Na** | 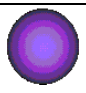 | **26.107** (9.06) | **77.494** (8.17) |
| **ADM•Na** (Orientation 1) | 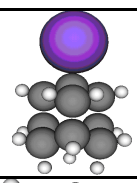 | **18.078** (9.06) | **27.790** (8.36) |
| **ADM•Na** (Orientation 2) | 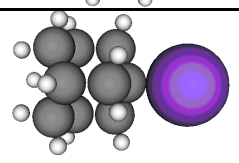 | **44.54**(9.90) | **53.60** (9.98) |
| **DIM•Na** (Orientation 1) | 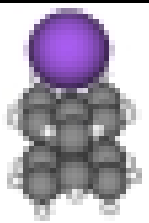 | **1.37** (7.67) | **103.41** (8.48) |
| **DIM•Na** (Orientation 2) | 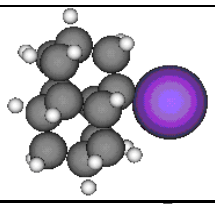 | **44.32** (11.06) | **54.24** (10.95) |
| **DIM•Na** (Orientation 3) | 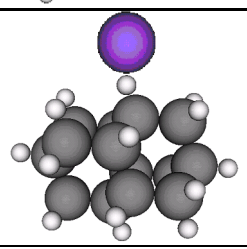 | **0.578** (10.63) | **23.654** (9.47) |

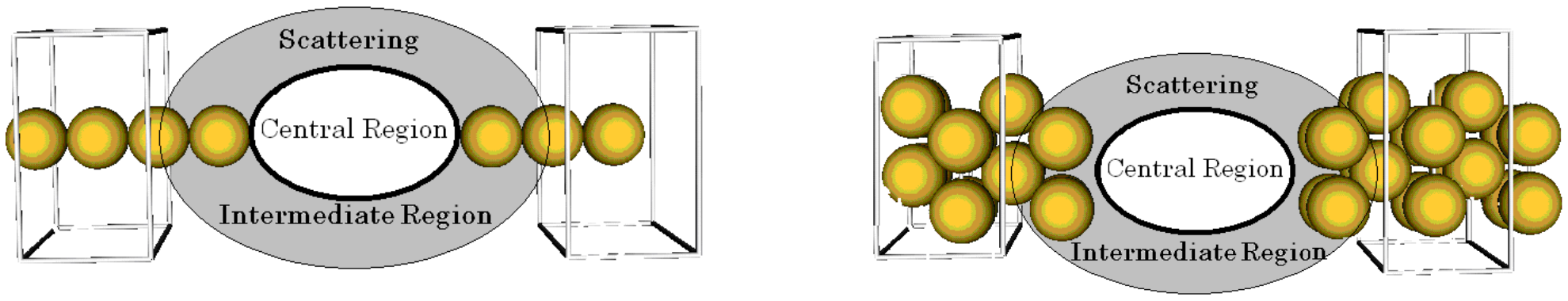

**FIGURE 1 -** The two semi-infinite linear and 2×2 gold (Au) chain electrodes

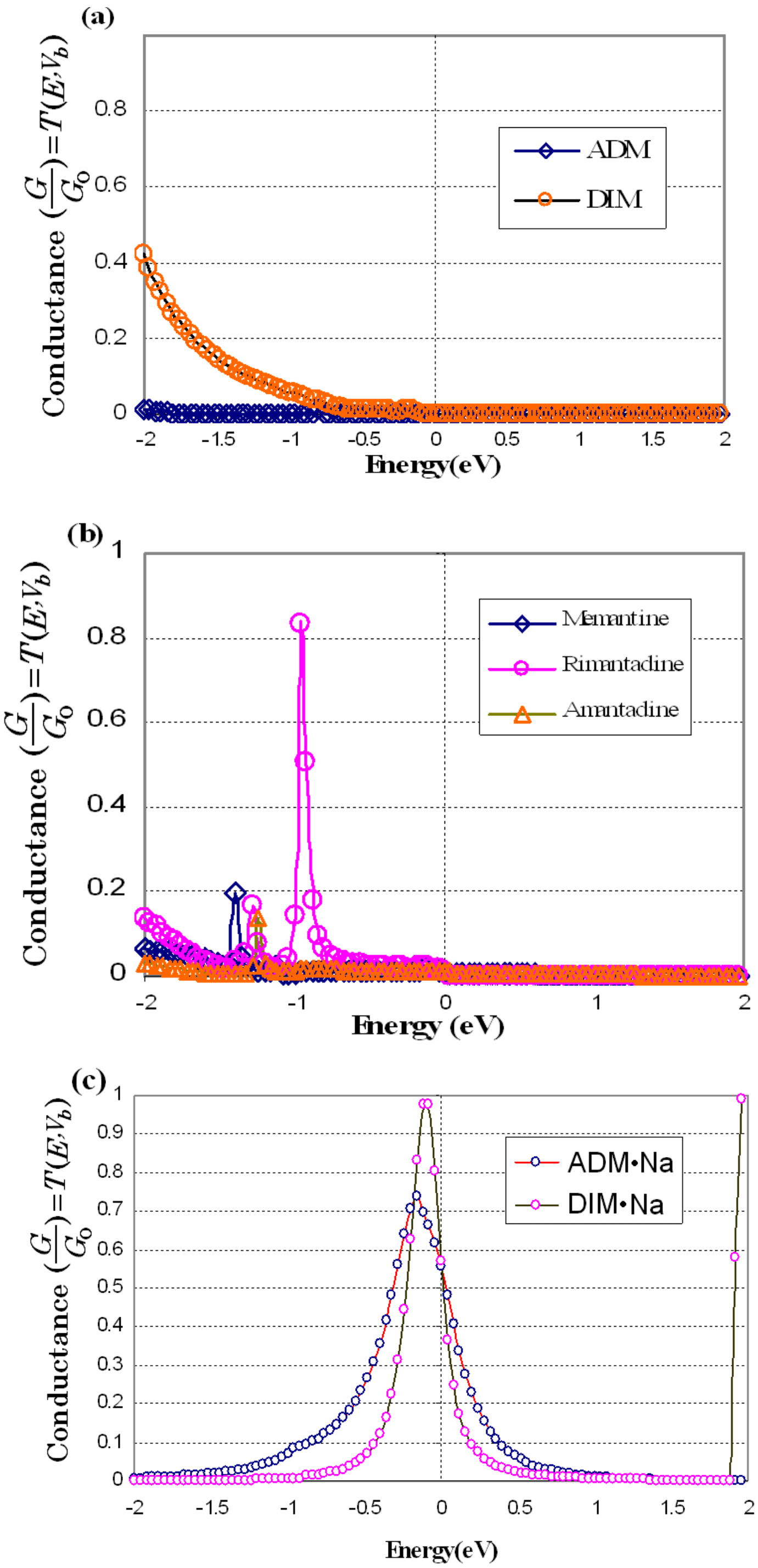

**FIGURE 2 -** Transmission spectrum in Au Linear electrodes cases. (a)Transmission spectrum of Group 1. (b)Transmission spectrum of Group 2 corresponding to "Orientation 1" in Table III. (c) Transmission spectrum of Group 3 corresponding to "Orientation 1" in Table III.

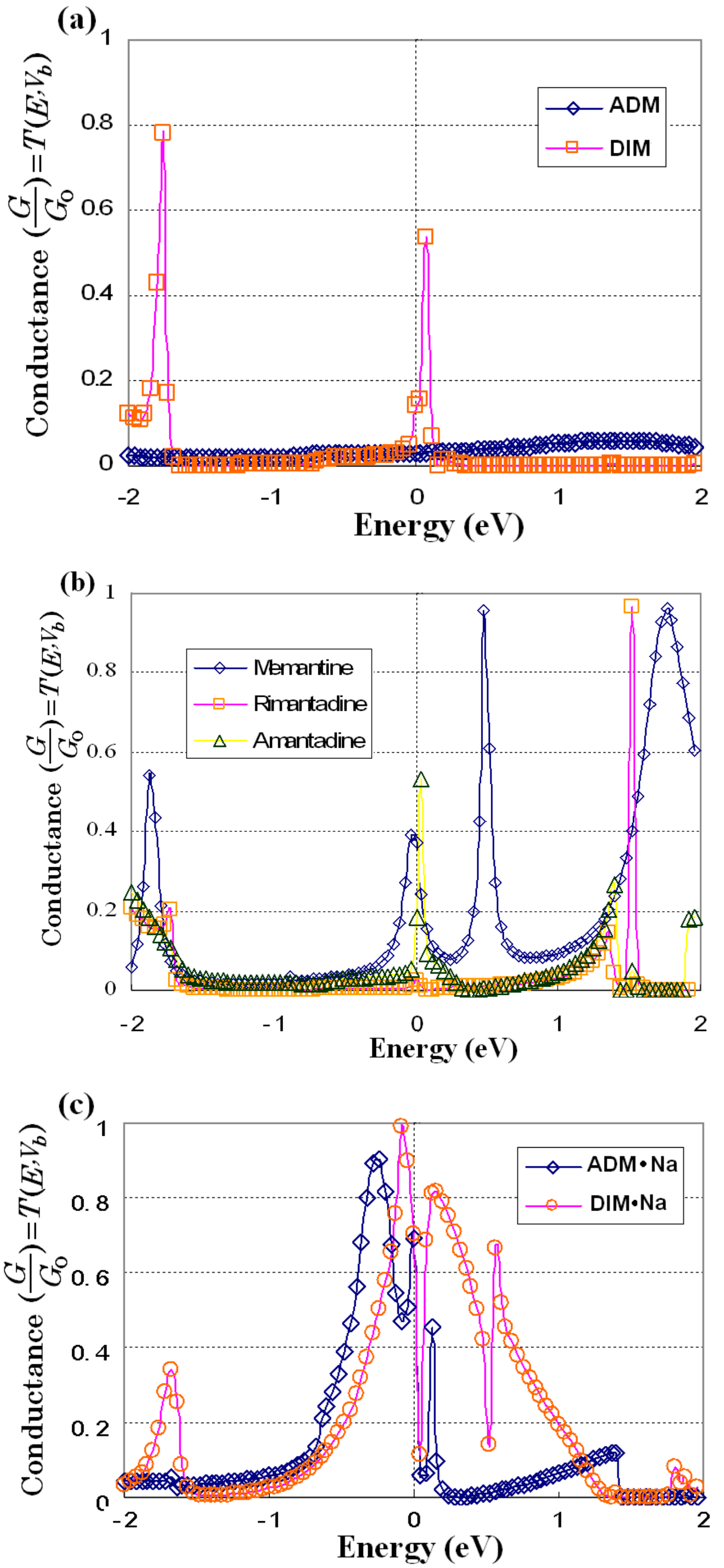

**FIGURE 3 -** Transmission spectrum in Au 2x2 unit cell cases. (a)Transmission spectrum of Group 1. (b) Transmission spectrum of Group 2 corresponding to "Orientation 1" in Table III. (c) Transmission spectrum of Group 3 corresponding to "Orientation 1" in Table III.

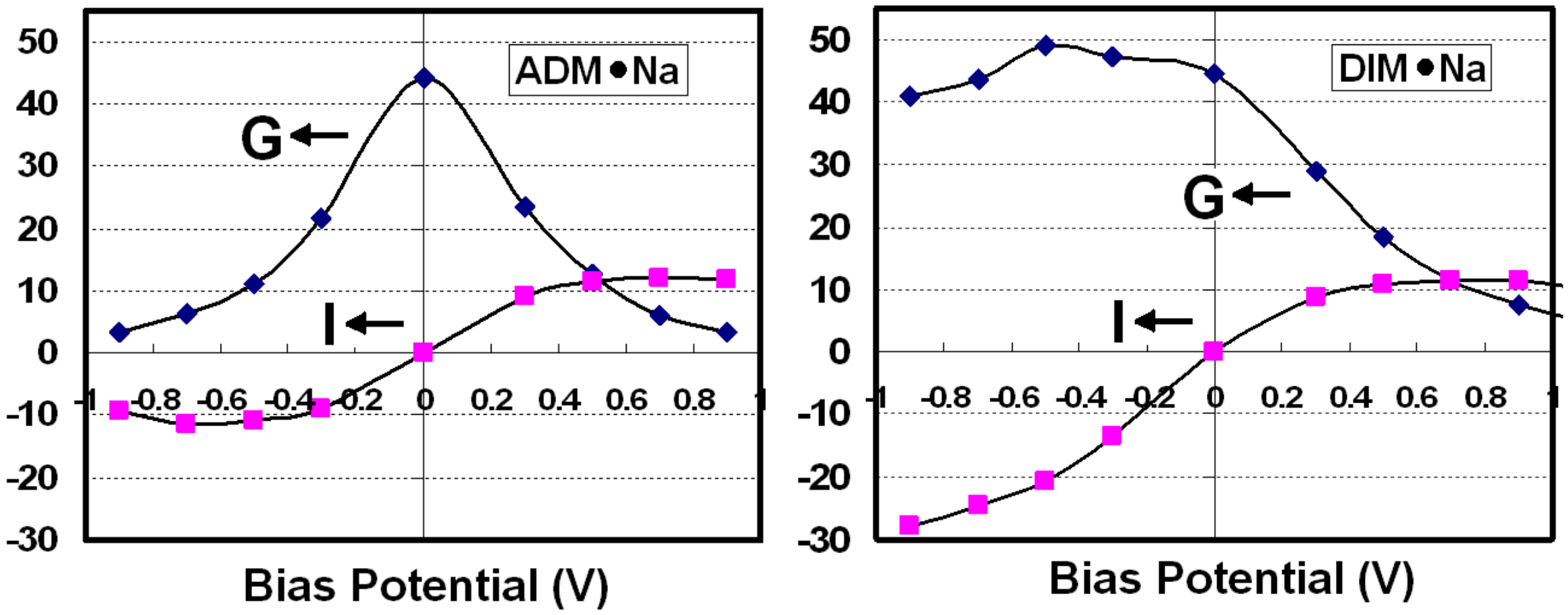

**FIGURE 4 – G** [*μS*] **vs.** *V* [*V*] and *I* [*μA*] vs. *V* [*V*] characteristics of ADM•Na and DIM•Na in linear electrodes corresponding to their horizontal orientations (Orientations 2 in Table III).